\documentclass[aps,prc,twocolumn,showpacs,superscriptaddress,floatfix,nofootinbib]{revtex4}

\usepackage{graphicx}
\usepackage{epstopdf}
\usepackage{dcolumn}
\usepackage{amsfonts}
\usepackage{braket}
\usepackage{amssymb,amsmath}

\begin{document}

\title{Time-dependent coupled-cluster method for atomic nuclei}

\author{D. A. Pigg} 

\affiliation{Department of Physics and Astronomy, Vanderbilt
  University, Nashville, Tennessee 37235, USA}

\affiliation{Physics Division, Oak Ridge National Laboratory, Oak
  Ridge, Tennessee 37831, USA}

\author{G. Hagen}

\affiliation{Physics Division, Oak Ridge National Laboratory, Oak
  Ridge, Tennessee 37831, USA}

\affiliation{Department of Physics and Astronomy, University of
  Tennessee, Knoxville, Tennessee 37996, USA} 

\author{H. Nam} 

\affiliation{National Center for Computational Sciences Division, Oak
  Ridge Leadership Computing Facility, Oak Ridge National Laboratory,
  P.O. Box 2008, Oak Ridge, Tennessee 37831, USA}

\author{T. Papenbrock} 

\affiliation{Department of Physics and Astronomy, University of
  Tennessee, Knoxville, Tennessee 37996, USA} 

\affiliation{Physics Division, Oak Ridge National Laboratory, Oak
  Ridge, Tennessee 37831, USA}

\date{\today}

\begin{abstract}
  We study time-dependent coupled-cluster theory in the framework of
  nuclear physics. Based on Kvaal's bi-variational formulation of this
  method [S. Kvaal, arXiv:1201.5548], we explicitly demonstrate that
  observables that commute with the Hamiltonian are conserved under
  time evolution. We explore the role of the energy and of the
  similarity-transformed Hamiltonian under real and imaginary time
  evolution and relate the latter to similarity renormalization group
  transformations. Proof-of-principle computations of $^4$He and
  $^{16}$O in small model spaces, and computations of the Lipkin model
  illustrate the capabilities of the method.
\end{abstract}


\maketitle

\section{Introduction}

Coupled-cluster (CC) theory was introduced in nuclear physics by
Coester and K{\"u}mmel \cite{Coe58,CoeKum60} more than 50 years ago,
and Cizek and Paldus \cite{Ciz66,Ciz69,PalLi99} further developed the
method for applications in quantum chemistry.  For reviews of this
method we refer the reader to
Refs.~\cite{Bis91,CraSch00,BarMus07,ShaBar09}. The popularity of
coupled-cluster theory is due to the attractive compromise the method
offers between accuracy on the one hand and computational cost on the
other hand. In nuclear structure and reactions, for instance, the
method has extended the reach of {\it ab initio} computations from
light $p$ shell nuclei~\cite{navratil2009} to medium-mass
nuclei~\cite{hagen2008,hagen2010b}.

Time-dependent coupled-cluster (TDCC) theory dates back more than 30
years. It was proposed by Hoodbhoy and
Negele~\cite{HooNeg78,HooNeg79}, with the aim to describe nuclear
collisions, and by Sch{\"o}nhammer and Gunnarson~\cite{SchGun78} to
compute spectral functions. Until now, however, nuclear collision
processes are usually described within time-dependent mean-field
methods~\cite{Dir30,bonche76,BriGiaEA76,UmaObe07,MeyGatEA09}.
Likewise, applications of TDCC in quantum chemistry have only been
sporadic~\cite{DalMon83,SekBar84,Seb85,Mon87,KocJor90,LatPra96,Pra02,HubKla11}.
For small amplitude oscillations, time-dependent coupled-cluster
theory leads to the computation of excited states within linear
response theory, and the method is routinely
used~\cite{DalMon83,TakPal86,KocJor90} in this framework. However, the
situation is different for large-amplitude oscillations.  We believe
that this lack of popularity is mainly due to conceptual problems
regarding the conservation of energy, and keeping the energy a real
(and not complex) number~\cite{HubKla11}. These aspects have only been
clarified very recently in Kvaal's formulation~\cite{Kva12} of TDCC as
a bi-variational theory~\cite{arponen83}, and the result is a real
energy that is conserved under time evolution. This approach opens the
way to many exciting applications. In this work, we present a formal
proof regarding the conservation of energy within Kvaal's formulation
of TDCC and also show that observables that commute with the
Hamiltonian are conserved under time evolution. We study in detail the
role of the similarity-transformed Hamiltonian under real and
imaginary time evolution and illustrate our results in applications to
the nuclei $^4$He and $^{16}$O in small model spaces and to the Lipkin
model.

This paper is organized as follows. In Sect.~\ref{sec:tdcc}, we
re-derive TDCC, set it in relation to time-independent CC, and discuss
formal aspects about conserved quantities.  In Sect.~\ref{sec:energy}
we focus on the energy. We explicitly demonstrate that the numerical
time-evolution of the bi-variational CC theory preserves the energy.
We also study the real and imaginary time-evolution of the
similarity-transformed Hamiltonian.  In Sect.~\ref{sec:apps} we
discuss applications of the method.  First, we show that the Fourier
analysis of the time-evolved coupled-cluster amplitudes of the nuclei
$^4$He and $^{16}$O yields excitation spectra. Second, we study the
Lipkin model and compare our results for observables with exact
results.  We finish with our conclusions.  Technical aspects are
relegated to the Appendix.

\section{\label{sec:tdcc}Coupled-cluster theory}
In this Section, we briefly re-derive the equations that govern TDCC
theory and prove that observables that commute with the Hamiltonian
are conserved under time evolution. We also consider TDCC in the limit
of linear response theory, and the limit of vanishing time dependence.

\subsection{\label{sec:tdccsd}Time-dependent coupled-cluster
  equations}
We are interested in solving the time-dependent Schr{\"o}dinger
equation 
\begin{eqnarray}
H(t)\Ket{\Psi(t)}&=&i\hbar\partial_t\Ket{\Psi(t)}.\label{eq:schro}
\end{eqnarray}
Here $H(t)$ is a time-dependent Hamiltonian.  In TDCC theory, the
many-body wave function results from an exponential excitation operator
acting on a product state
\begin{eqnarray}
\Ket{\Psi}=e^{S(t)}\Ket{\Phi} \ .\label{eq:expans}
\end{eqnarray}
Here
\begin{equation}
\Ket{\Phi}=\prod_{p=1}^Aa_p^{\dag}(t)\Ket{0}
\label{eq:tdccphi}
\end{equation}
is an $A$-body product state that serves as the reference state. The
creation operator $a_p^{\dag}$ creates a fermion in the
single-particle state $p$, while an annihilation operator $a_p$ would
remove a fermion from the single-particle state $p$.  The creation and
annihilation operators are allowed to depend on time, and they fulfill
the usual anti-commutation relations at equal times.  The time
dependence of the creation and annihilation operators is~\cite{HooNeg79}
\begin{equation}
\label{adot}
\partial_t a_q^\dagger(t)= \sum_p \langle p(t)|\dot{q}(t)\rangle a^\dagger_p(t) .
\end{equation}
Here $|q(t)\rangle=a^\dagger_q(t)|0\rangle$ and $|\dot{q}(t)\rangle$
denotes the time derivative of $|q(t)\rangle$.  To simplify notation,
we will suppress the explicit time argument and denote $a_q(t)$ as
$a_q$ and $|q(t)\rangle$ as $|q\rangle$. In Eq.~(\ref{eq:expans}) the
cluster operator $S(t)$ is
\begin{eqnarray}
S(t)=s_0(t)+S_1(t)+S_2(t)+S_3(t)+\textellipsis+S_A(t) \ .
\label{eq:ssum}
\end{eqnarray}
Here $s_0(t)$ is a complex phase while the operators $S_1(t),\ldots, S_A(t)$
generate 1$p$-1$h$, 2$p$-2$h$, 3$p$-3$h$, and up to
$Ap$-$Ah$ excitations within the reference state $\Ket{\Phi}$, with
\begin{eqnarray}
\lefteqn{S_n(t)=}\\
&&\frac{1}{(n!)^2}\sum_{i_1...i_n,a_1...a_n}s_{i_1...i_n}^{a_1...a_n}(t)
a_{a_1}^{\dag}\textellipsis a_{a_n}^{\dag}a_{i_n}\textellipsis
a_{i_1} \nonumber \ .
\label{eq:sgen}
\end{eqnarray}
The functions $s_0(t)$ and $s_{i_1...i_n}^{a_1...a_n}(t)$ are the
unknown time-dependent excitation amplitudes.  In what follows, we
also suppress the explicit time argument of the
$n$p-$n$h cluster operators $S_n(t)$, the phase $s_0(t)$, and the
excitation amplitudes $s_{i_1...i_n}^{a_1...a_n}(t)$.  In addition, we
will use the indices $i,j,k,...$ and $a,b,c,...$ to label occupied and
unoccupied states in the reference~(\ref{eq:tdccphi}), respectively,
and the indices $p,q,r,...$ to label any state.

We are interested in the time evolution of the cluster operator $S$.
For computational efficiency, we employ the coupled-cluster with
singles-and-doubles (CCSD) approximation and truncate $S$
at the 2$p$-2$h$ excitation level
\begin{eqnarray}
  S&=&s_0+S_1+S_2\nonumber\\
  &=&s_0+\sum_{ia}s_i^aa_a^{\dag}a_i+\frac{1}{4}
\sum_{ijab}s_{ij}^{ab}a_a^{\dag}a_b^{\dag}a_ja_i \ .
\label{eq:soft1}
\end{eqnarray}

To derive the equations of motion for the amplitudes $s_0$, $s_i^a$,
and $s_{ij}^{ab}$ we left-multiply the time-dependent Schr{\"o}dinger
equation~(\ref{eq:schro}) with the operator $e^{-S}$, such that
\begin{equation}
\overline{H}|\Phi\rangle =
i\hbar e^{-S}\partial_t e^S|\Phi\rangle\ .
\label{td}
\end{equation}
Here, we introduced the similarity-transformed Hamiltonian
\begin{equation}
\label{sim}
\overline{H}\equiv e^{-S}He^S
\end{equation}
For the computation of the similarity transformation
$\overline{B}\equiv e^{-S}B e^S$ of an operator $B$ we use the
Baker-Campbell-Hausdorff expansion,
\begin{eqnarray}
e^{-S}Be^{S}&=&B+[B,S]+\frac{1}{2!}[[B,S],S]\label{eq:cche}\\
&+&~\frac{1}{3!}[[[B,S],S],S]\nonumber\\
&+&~\frac{1}{4!}[[[[B,S],S],S],S]+\textellipsis\nonumber \ .
\end{eqnarray}
For a two-body operator, this expansion truncates naturally
at its fourth nested commutator.  For $B=\partial_t$, the expansion in
Eq.(\ref{eq:cche}) reduces to
\begin{eqnarray}
e^{-S}\partial_te^{S}&=&\partial_t+\dot{S}+\frac{1}{2}[\dot{S},S]\label{eq:hbctdccs}
\end{eqnarray}
for any truncation of $S$. Note that the commutator $[\dot{S},S]$ is
only nonzero if the single-particle states depend explicitly on time
[see Eq.~(\ref{adot})].

We insert the CCSD ansatz~(\ref{eq:soft1}) into Eq.~(\ref{td});
left-project with the Hermitian adjoints of the states $|\Phi\rangle$,
$|\Phi_i^a\rangle\equiv a^\dagger_aa_i|\Phi\rangle$, and
$|\Phi_{ij}^{ab}\rangle\equiv
a^\dagger_aa^\dagger_ba_ja_i|\Phi\rangle$; and then use
Eq.~(\ref{eq:hbctdccs}).  In what follows, we limit ourselves to
time-independent basis functions. The equations of motion for the $S$
amplitudes within a time-dependent basis are presented in
Appendix~\ref{app:tdb}. We obtain 
\begin{eqnarray}
i\hbar\dot{s}_0&=&\Bra{\Phi}\overline{H}\Ket{\Phi} , 
\label{eq:tdcce2}\\
i\hbar\dot{s}_i^a &=&\Bra{\Phi_{i}^{a}}\overline{H}\Ket{\Phi} , 
\label{eq:tdccs12}\\
i\hbar\dot{s}_{ij}^{ab} &=&\Bra{\Phi_{ij}^{ab}}\overline{H}\Ket{\Phi} .
\label{eq:tdccs22}
\end{eqnarray}
The left-hand sides of these equations are matrix elements of the
similarity-transformed Hamiltonian $\overline{H}\equiv e^{-S}He^{S}$.
The corresponding algebraic expressions can be worked out
diagrammatically and are well known from time-independent
CCSD~\cite{CraSch00,ShaBar09}. The resulting expressions are linear in
the matrix elements of the Hamiltonian and nonlinear in the cluster
amplitudes. It is useful to rewrite the equations~(\ref{eq:tdcce2}) to
(\ref{eq:tdccs22}) in operator form as
\begin{equation}
i\hbar \dot{S} = \overline{H}_1 .
\label{tibasis}
\end{equation}
Here $\overline{H}_1$ denotes the operator that is obtained from the
similarity-transformed Hamiltonian $\overline{H}$ when all but the
matrix elements on the right hand side of
Eqs.~(\ref{eq:tdcce2}-\ref{eq:tdccs22}) are set to zero. When viewing
$\overline{H}$ as a matrix (See Eq.~(\ref{eq:hbarblock}) below), the first
column of the matrix of $\overline{H}_1$ is thus identical to the
first column of $\overline{H}$, while all other columns are zero.

For the computation of observables within TDCC we follow
Ref.~\cite{Kva12}. Coupled-cluster theory employs a non-Hermitian
Hamiltonian $\overline{H}$ and must thus be formulated in a
bi-variational approach, i.e. the left (bra) and
right (ket) states must be varied independently. In this formulation,
the energy expectation value is
\begin{eqnarray}
E&=&\Bra{\Phi}L\overline{H}\Ket{\Phi}\label{eq:ccfunctional}. 
\end{eqnarray}
Here $L$ is a linear de-excitation operator
\begin{eqnarray}
 L=l_0+\sum_{ia}l_a^ia_i^{\dag}a_a+\frac{1}{4}
\sum_{ijab}l_{ab}^{ij}a_i^{\dag}a_j^{\dag}a_ba_a,
\label{eq:ldeex}
\end{eqnarray}
and $l_0$, $l_a^i$, and $l_{ab}^{ij}$ are the amplitudes associated
with up to 2$h$-2$p$ de-excitations~\cite{ShaBar09}.  From
Eq.~(\ref{eq:ccfunctional}), it is clear that the left many-body state
is given by
\begin{eqnarray}
\bra{\Psi_L}&\equiv&\bra{\Phi}Le^{-S}\label{eq:leftstate}.
\end{eqnarray}
To obtain the time evolution of the operator $L$, we insert
$\bra{\Psi_L}$ of Eq.~(\ref{eq:leftstate}) into the time-dependent Schr{\"o}dinger
equation, right-multiply by $e^S$ and obtain 
\begin{eqnarray}
\label{left}
-i\hbar\dot{l}_0&=&0 , \\
-i\hbar\dot{l}_a^i&=&\langle\Phi|L\overline{H}|\Phi_i^a\rangle , \\
-i\hbar\dot{l}_{ab}^{ij}&=&\langle\Phi|L\overline{H}|\Phi_{ij}^{ab}\rangle . 
\end{eqnarray}
Again, we can write these equations in operator form as 
\begin{eqnarray}
-i\hbar\dot{L}=L\overline{H}-L\overline{H}_1\label{eq:levo1},
\end{eqnarray}
and here it is understood that the right hand side, when written in
matrix form of Eq.~(\ref{eq:hbarblock}), has nonzero matrix elements
in all but the first row.  Note that $l_0=1$ for proper normalization.

Let us explicitly check that the energy~(\ref{eq:ccfunctional}) is conserved. 
From the definition $\overline{H}=e^{-S}He^S$ and Eq.~(\ref{tibasis}) we find
\begin{eqnarray}
\label{doth}
\dot{\overline{H}}&=&[\overline{H},\dot{S}] =-{i\over\hbar}[\overline{H},\overline{H}_1] . 
\end{eqnarray}
Thus
\begin{eqnarray}
\label{econst}
\dot{E}&=&\bra{\Phi}\dot{L}\overline{H}\ket{\Phi}+\bra{\Phi}L\dot{\overline{H}}\ket{\Phi}\nonumber\\
&=&\frac{i}{\hbar}\Big(\bra{\Phi}(L\overline{H}-L\overline{H}_1)\overline{H}\ket{\Phi}\nonumber\\
&&+\bra{\Phi}L\overline{H}_1\overline{H}\ket{\Phi}-\bra{\Phi}L\overline{H}\,\overline{H}_1\ket{\Phi}\Big)\nonumber\\
&=&0 . 
\end{eqnarray}
Here we used Eq.~(\ref{eq:levo1}) and
$\overline{H}\ket{\Phi}=\overline{H}_1\ket{\Phi}$. 

We can extend these results to any observable $B$ that commutes
with the Hamiltonian. We have for the expectation value
$\langle B\rangle \equiv \bra{\Phi}L\overline{B}\ket{\Phi}$ 
\begin{eqnarray}
\dot{\langle B\rangle}&=&\bra{\Phi}\dot{L}\overline{B}\ket{\Phi}+\bra{\Phi}L\dot{\overline{B}}\ket{\Phi}\nonumber\\
&=& {i\over\hbar}\bra{\Phi} L[\overline{H},\overline{B}]\ket{\Phi}\nonumber\\
&=& 0 .
\end{eqnarray}
Here, we used
$\dot{\overline{B}}=(i/\hbar)[\overline{H},\overline{B}]$ as the
generalization of Eq.~(\ref{doth}), and
$[\overline{H},\overline{B}]=0$ follows from $[H,B]=0$. Note that this
proof addresses a long-standing problem already raised by
Monkhorst~\cite{Mon87}.

For the computation of observables, one employs the normal-ordered
one-body and two-body density matrices
\begin{eqnarray}
({\rho_{qp}^n})_N&=&\Bra{\Phi}Le^{-S}\{p^{\dag}q\}e^{S}\Ket{\Phi}_C , \label{eq:3den1}\\
({\rho_{rspq}^n})_N&=&\Bra{\Phi}Le^{-S}\{p^{\dag}q^{\dag}sr\}e^{S}\Ket{\Phi}_C  . \label{eq:3den2}
\end{eqnarray}
Here, only connected ($C$) diagrams contribute.  Due to the linearity
of $L$, these densities have finite expansions.

\subsection{\label{sec:tdtocc}Time-independent coupled-cluster equations}
Let us re-derive the well-known time-independent CCSD equations from
TDCC theory by requiring that the cluster operators $S_1$ and $S_2$,
and the single-particle basis do not depend on time. Thus we obtain
from the Eqs.~(\ref{eq:tdccs12}) and (\ref{eq:tdccs22})
\begin{eqnarray}
\Bra{\Phi_i^a}\overline{H}\Ket{\Phi}&=&0 \ ,
\label{eq:tia}\\
\Bra{\Phi_{ij}^{ab}}\overline{H}\Ket{\Phi}&=&0 \ .
\label{eq:tijab}
\end{eqnarray}
These are the time-independent CCSD equations. They state that the
reference state $|\Phi\rangle$ is an eigenstate of the
similarity-transformed Hamiltonian $\overline{H}$ in the space of
1$p$-1$h$ and 2$p$-2$h$ excitations. Thus the corresponding eigenvalue is
the energy
\begin{eqnarray}
E_0=\Bra{\Phi}\overline{H}\Ket{\Phi} \ ,
\label{eq:eticc}
\end{eqnarray}
and Eq.~(\ref{eq:tdcce2}) yields
\begin{eqnarray}
s_0=-\frac{i}{\hbar}Et \
\label{eq:defs0}
\end{eqnarray}
up to an irrelevant constant.

\subsection{\label{sec:ese}Coupled-cluster equations of motion}
The equation-of-motion CC methods~\cite{KocJor90,StaBar93} are commonly used to
compute excited states within the CC method. For
completeness, we briefly re-derive the corresponding equations 
from the time-dependent formalism.  Again, we assume that the Hamiltonian and
the single-particle basis do not explicitly depend on time.  For the
solutions of Eqs.~(\ref{eq:tdccs12}) and (\ref{eq:tdccs22}), we make
the ansatz
\begin{eqnarray}
s_i^a(t) &=& t_i^a +r_i^a(t) , \nonumber\\
s_{ij}^{ab}(t) &=& t_{ij}^{ab} +r_{ij}^{ab}(t) , 
\label{ansatz}
\end{eqnarray}
assume that the time-independent
amplitudes $t_i^a$ and $t_{ij}^{ab}$ fulfill the time-independent CCSD
equations~(\ref{eq:tia},\ref{eq:tijab}), and that the remainders
$r_i^a$ and $r_{ij}^{ab}$ are small perturbations, i.e.  $S=T+R$ with
$|R|\ll |T|$. In leading order we thus have
\begin{equation}
\label{lo}
\overline{H}\approx e^{-T}He^T \equiv\overline{H}_T.
\end{equation}
We insert $S=T+R$ into Eq.~(\ref{tibasis}), employ
Eqs.~(\ref{eq:tia}), (\ref{eq:tijab}), and the
expansion~(\ref{eq:cche}), and keep only the terms first-order in $R$.
This yields the operator equation
\begin{equation}
i\hbar \dot{R} = \overline{H}_T+ [\overline{H}_T,R] \ .
\end{equation}
Then, assuming, a simple harmonic time-dependence,
\begin{eqnarray}
r_i^a(t)&=& r_i^a e^{-i\omega t}\\
r_{ij}^{ab}(t)&=& r_{ij}^{ab} e^{-i\omega t},
\end{eqnarray}
we find that the amplitudes and frequencies solve the eigenvalue problem
\begin{eqnarray}
\hbar\omega r_i^a = \langle\Phi_i^a|[\overline{H}_T,R]|\Phi\rangle\ ,\nonumber\\
\hbar\omega r_{ij}^{ab} = \langle\Phi_{ij}^{ab}|[\overline{H}_T,R]|\Phi\rangle.\label{eq:eom}
\end{eqnarray}
These are the well known equations of motion for the computation of
excited states within CCSD. 

It is also interesting to study the left equation~(\ref{eq:levo1}) for
$S=T+R$, with $|R|,|L|\ll T$. Keeping terms linear in $R$ and $L$
yields in leading order
\begin{eqnarray}
-i\hbar\dot{L}=L(\overline{H}_T-E_0)
\end{eqnarray}
since $(\overline{H}_T)_1=E_0$. Again we assume a harmonic time-dependence
\begin{eqnarray}
l^i_a(t)&=& l^i_a e^{i\omega t} , \\
l^{ij}_{ab}(t)&=& l^{ij}_{ab} e^{i\omega t},
\end{eqnarray}
and find the well-known left eigenvalue problem 
\begin{eqnarray}
\label{lefteom}
L\overline{H}_T=(E_0+\omega) L .
\end{eqnarray}

\section{\label{sec:energy}Role of energy}

In this Section, we focus on the role of the energy in TDCC theory.
First, we probe the conservation of energy~(\ref{econst}) numerically
for simple models of light nuclei.  Next, we study the time-dependent
eigenvalues of the similarity-transformed Hamiltonian $\overline{H}$
and find that these cannot be viewed as energies. This opens the way
to employ imaginary-time evolutions of the similarity-transformed
Hamiltonian for the computation of the ground-state energy.  In what
follows we employ a low-momentum two-body interaction produced by a
similarity renormalization group transformation~\cite{BogFurEA10} of
the N$^3$LO interaction from chiral effective field theory~\cite{EM}
at the cutoff $\Lambda=1.9$~fm$^{-1}$. We use a small model space of
four oscillator shells.  For the time integration of $S$ and $L$
according to Eqs.~(\ref{tibasis}) and (\ref{eq:levo1}), we employ the
fourth-order Runge-Kutta method.

\subsection{\label{sec:realteng}Time evolution of the energy functional}

\begin{figure}[th]
\begin{center}
\includegraphics[keepaspectratio=true,scale=0.300]{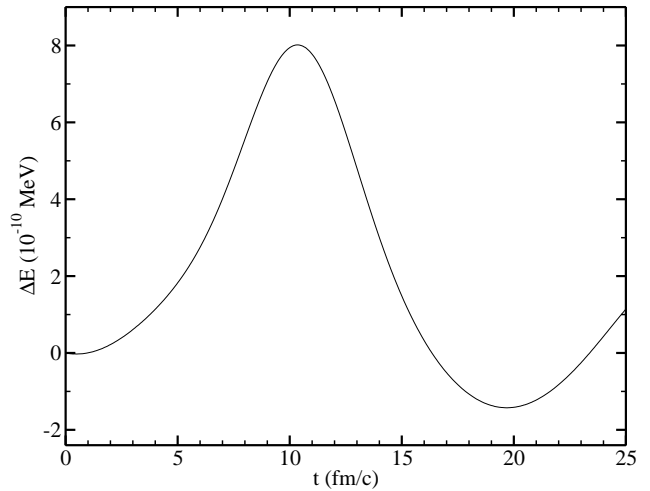}
\end{center}
\caption{The time evolution of the change in total energy of $^{4}$He
  from its initial value.}\label{fig:he4deltae}
\end{figure}

For the time evolution, we iteratively solve Eqs.~(\ref{tibasis}) and
(\ref{eq:levo1}). We set $l_0=1$ and neglect the complex phase $s_0$,
and periodically compute the difference between the total energy,
according to Eq.~(\ref{eq:ccfunctional}), and its initial value.
Figure~\ref{fig:he4deltae} shows this difference in total binding energy
as a function of time for the $^{4}$He nucleus with a time step of
width 0.05~fm/$c$. The change in energy is very small throughout the
time-evolution and energy is conserved for all practical purposes.  In
practice, the precision to which the method conserves energy has a
notable dependence on the time step width: we obtain changes to energy
on the order of 10$^{-10}$ MeV using a step width of 0.05~fm/$c$ and
changes to energy of average order $10^{-2}$~MeV for a step width of
1~fm/$c$.

\subsection{\label{sec:realthbar}Time evolution of $\overline{H}$}

We are interested in the eigenvalues of the similarity-transformed
Hamiltonian~(\ref{sim}). For a time-independent Hamiltonian $H$, the
eigenvalues of $\overline{H}$ do not depend on time if the similarity
transformation is carried out exactly. For an $A$-body system, a
complete model space consists of up to $Ap$-$Ah$ excitations of the
reference state, but this is computationally not feasible in
general. Instead, one chooses a model space of up to $2p$-$2h$
excitations within the CCSD approximation. This truncation renders the
similarity transformation inexact and it is thus no surprise that the
eigenvalues of $\overline{H}$ are not constant in time.

To illustrate the effects of the truncations, we consider three
distinct cases.  (\textit{i}) In the case ``TD-CCSD (2$p$-2$h$),'' we
employ $S=S_1+S_2$ in the computation of $\overline{H}$ [i.e. we time
evolve the $S$ amplitudes according to Eqs.~(\ref{eq:tdccs12}) and
(\ref{eq:tdccs22})], and compute the lowest eigenvalue of
$\overline{H}$ within a basis including up to 2$p$-2$h$ excited
states. Clearly, this case has truncations in the model space for
nuclei with $A>2$.  (\textit{ii}) In the case ``TD-CCS (2$p$-2$h$),''
we employ $S=S_1$ [i.e. we set $S_2=0$, solve the time-dependent
coupled-cluster singles (CCS) equation~(\ref{eq:tdccs12})]. We compute
the ground-state energy within a basis including up to 2$p$-2$h$
excited states. Here, the similarity-transformation is simpler than in
case ({\it i}), but the model space is incomplete for nuclei with
$A>2$.  (\textit{iii}) In the case ``TD-CCS (3$p$-3$h$),'' we again
solve the time-dependent CCS equation~(\ref{eq:tdccs12}), but compute
the ground-state energy within a basis including up to 3$p$-3$h$
excited states $|\Phi_{ijk}^{abc}\rangle$. Here, no approximation is
made for a nucleus with $A=3$. In what follows we will consider these
distinct cases for nuclei with mass number $A=3$ and $A=4$.

\begin{figure}[h]
\begin{center}
\includegraphics[keepaspectratio=true,scale=0.300]{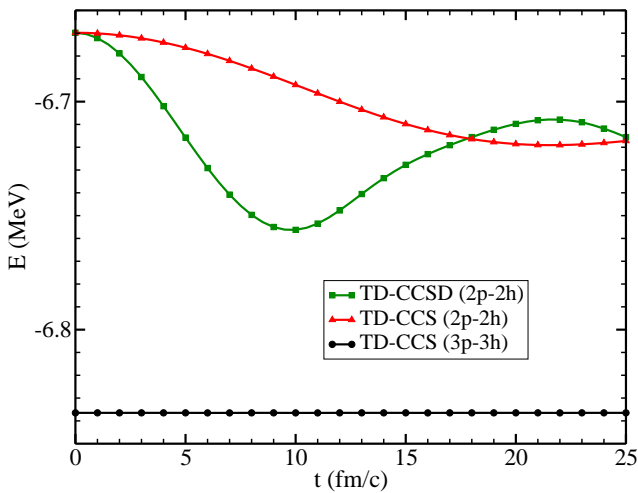}
\end{center}
\caption{(Color online) The time evolution of the lowest eigenvalue of the
  similarity-transformed Hamiltonian $\overline{H}$ for $^{3}$H
  computed for three distinct cases (see text for
  details).}\label{fig:h3en}
\end{figure}

Figure~\ref{fig:h3en} shows the eigenvalues of the
similarity-transformed Hamiltonian of $^3$H as a function of time for
the three cases discussed above.  As expected, the energy is only
conserved for case ({\it iii}). Case ({\it i}) and ({\it ii}) are
different similarity transformations and yield different
$\overline{H}$.

\begin{figure}[h]
\begin{center}
\includegraphics[keepaspectratio=true,scale=0.300]{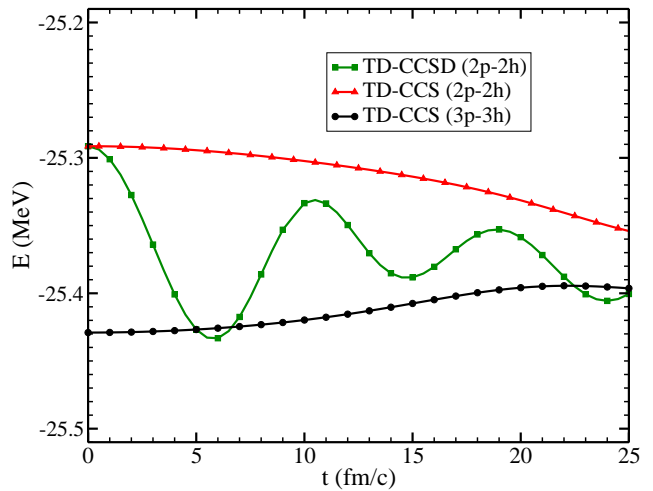}
\end{center}
\caption{(Color online) Same as Fig.~\ref{fig:h3en} but for
  $^{4}$He}\label{fig:he4en}
\end{figure}

Figure~\ref{fig:he4en} shows the eigenvalues of the
similarity-transformed Hamiltonian of $^4$He as a function of time for
the three cases discussed above.  As expected, the energy is not
conserved in any of the cases. The fluctuations of the energy are
smallest for case ({\it iii}) which exhibits the least truncation.

The results of this subsection explain clearly why matrix
elements~\cite{HubKla11} or eigenvalues of the similarity-transformed
Hamiltonian cannot serve as conserved energies. However, these results
also open up the possibility to employ imaginary-time evolution as a
projection technique.

\subsection{\label{sec:imagthbar}Imaginary-time evolution of $\overline{H}$}

Imaginary time propagation of a given quantum state projects out the
lowest-energy state that has a nonzero overlap with this state.
We set $t=-i\hbar\tau$ with real $\tau$ and obtain the evolution
equations
\begin{eqnarray}
\partial_\tau s_i^a &=& - \Bra{\Phi_{i}^{a}}e^{-S}He^S\Ket{\Phi}\label{eq:it2},\\
\partial_\tau s_{ij}^{ab}&=& -\Bra{\Phi_{ij}^{ab}}e^{-S}He^S\Ket{\Phi}\label{eq:it3}
\end{eqnarray}
from Eqs.~(\ref{eq:tdccs12}) and (\ref{eq:tdccs22}). It is useful to
consider the imaginary time evolution of the Hamiltonian directly. The
corresponding evolution equation is
\begin{equation}
\label{flow}
\partial_\tau \overline{H}(\tau) = [\overline{H}(\tau),\partial_\tau S] \ .
\end{equation}
Eq.~(\ref{flow}) shows that $\partial_\tau S$ generates the similarity
transformation of the Hamiltonian.  It is similar to the
similarity renormalization group equations employed in taming
interactions~\cite{GlaWil93,Weg94,BogFurEA07} and in computations of
finite nuclei~\cite{TsuBogEA12}. The evolution stops (i.e. it reaches
a fixed point) once the right hand side of Eq.~(\ref{flow}) vanishes.
At the fixed point the matrix elements of $\overline{H}$ fulfill the
time-independent CCSD Eqs.~(\ref{eq:tia},\ref{eq:tijab}). Thus, the
imaginary time evolution emerges as an alternative way of solving the
coupled-cluster equations.

\begin{figure}[h]
\begin{center}
\includegraphics[keepaspectratio=true,scale=0.325]{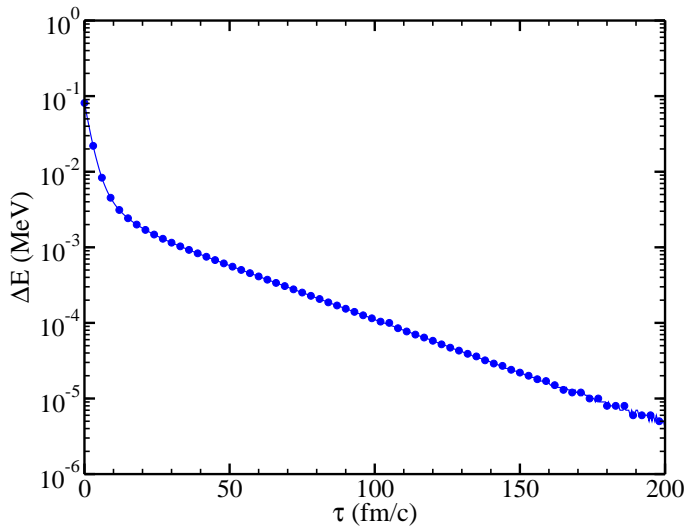}
\end{center}
\caption{(Color online) The magnitude of the difference between the total
  energy and CCSD ground-state energy for $^{4}$He as a function of
  imaginary time $\tau$.}\label{fig:he4itp}
\end{figure}

As an example, we consider the imaginary time evolution for $^4$He.
For the initial conditions we choose $s_i^a=0=s_{ij}^{ab}$.  We then
propagate $s_i^a$ and $s_{ij}^{ab}$ according to Eqs.~(\ref{eq:it2})
and (\ref{eq:it3}) while periodically computing the ground-state energy
as the right eigenvalue of $\overline{H}$. Figure~\ref{fig:he4itp}
shows the difference $\Delta E = |E(\tau)-E_{\rm CCSD}|$ as a function
of $\tau$, obtained using 400 steps of width 0.5~fm/$c$. Here, $E_{\rm
  CCSD}$ is the energy that one can obtain from solving the CCSD
equations~(\ref{eq:tia},\ref{eq:tijab}) directly.  The convergence to
the ground state energy is exponentially fast, and the exponent is
related to the energy gap to the first excited state.

\begin{figure}[t]
\begin{center}
\includegraphics[keepaspectratio=true,scale=0.300]{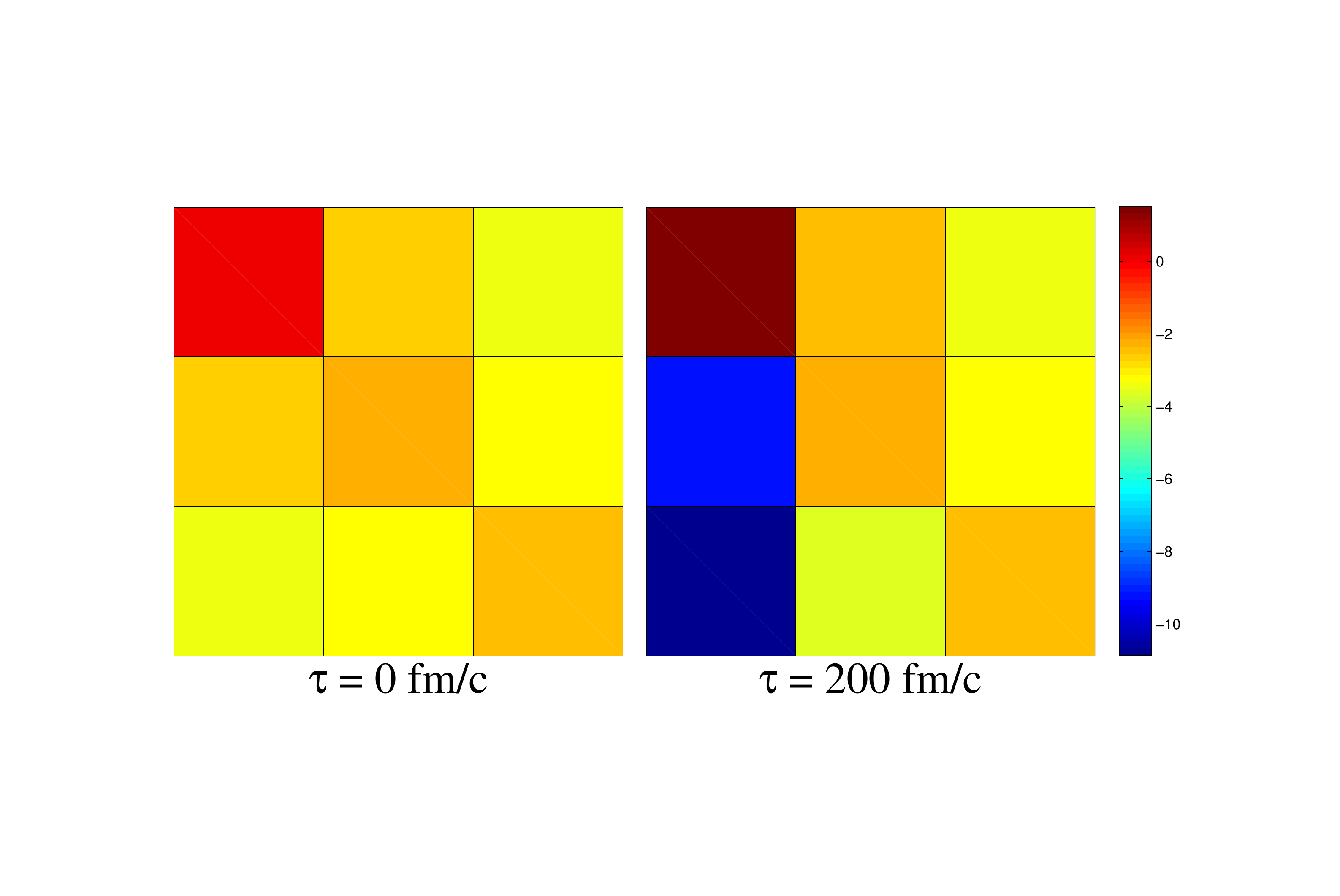}
\end{center}
\caption{(Color online) Logarithmic averages over the elements of
  $\overline{H}(\tau)$ for $^4$He at $\tau=$~0 fm/c and $\tau=$~200
  fm/c, where $\overline{H}(\tau)$ has the form given
  in~(\ref{eq:hbarblock}).}\label{fig:he4itc}
\end{figure}

Let us also visualize the evolution of the $\tau$-dependent
similarity-transformed Hamiltonian. We present the 
Hamiltonian in a block format as
\begin{equation}
\overline{H}=\begin{pmatrix}
\Bra{\Phi}\overline{H}\Ket{\Phi}
&\Bra{\Phi}\overline{H}\Ket{\Phi_k^c}
&\Bra{\Phi}\overline{H}\Ket{\Phi_{kl}^{cd}}\\\\
\Bra{\Phi_i^a}\overline{H}\Ket{\Phi}
&\Bra{\Phi_i^a}\overline{H}\Ket{\Phi_k^c}
&\Bra{\Phi_i^a}\overline{H}\Ket{\Phi_{kl}^{cd}}\\\\
\Bra{\Phi_{ij}^{ab}}\overline{H}\Ket{\Phi}
&\Bra{\Phi_{ij}^{ab}}\overline{H}\Ket{\Phi_k^c}
&\Bra{\Phi_{ij}^{ab}}\overline{H}\Ket{\Phi_{kl}^{cd}}
\label{eq:hbarblock}
\end{pmatrix}
\end{equation}
and plot logarithms of the averages of the magnitudes of all elements
within a given block.  Figure~\ref{fig:he4itc} shows contour plots of
these averages taken at $\tau=0$~fm/$c$ and $\tau=200$~fm/$c$, with a
step size of $\delta\tau=0.5$~fm/$c$. We see that the Hamiltonian is
driven to a form that makes the reference state $|\Phi\rangle$ a right
eigenstate and the element $\langle\Phi|\overline{H}|\Phi\rangle$ the
corresponding eigenvalue. At $\tau=200$~fm/$c$ the time-independent
CCSD Eqs.~(\ref{eq:tia}) and (\ref{eq:tijab})) have practically been solved.

\section{\label{sec:apps}Applications}

In this Section we study two applications of the TDCC method.  First,
we compute the spectra of excited state energies for $^4$He and
$^{16}$O as the Fourier transforms of the time-evolved cluster
amplitudes and compare the results to those obtained by solving the
time-independent coupled-cluster equations of motion.  Second, we
revisit the interacting Lipkin system and compare with the results by
Hoodbhoy and Negele~\cite{HooNeg78}.

\subsection{Energy spectra from Fourier transforms}\label{sec:ft}

\begin{figure}[h]
\begin{center}
\includegraphics[keepaspectratio=true,scale=0.300]{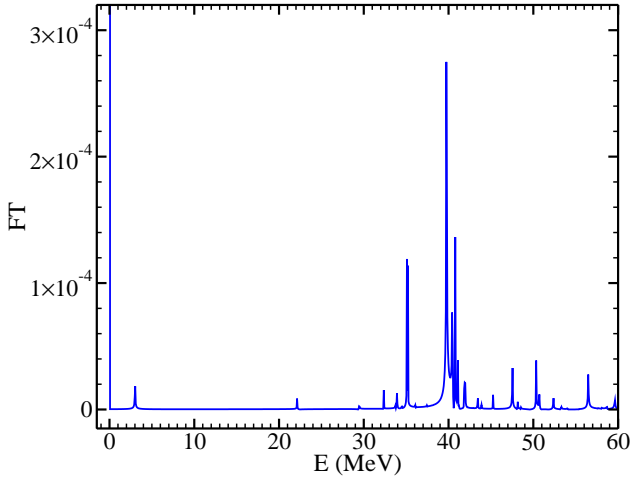}
\end{center}
\caption{(Color online) Fourier transform of a randomly-selected cluster amplitude
  $s_i^a$ or $s_{ij}^{ab}$ for $^4$He. Energies are measured relative
  to the ground-state energy.}\label{fig:he4ft}
\end{figure}

\begin{figure}[h]
\begin{center}
\includegraphics[keepaspectratio=true,scale=0.300]{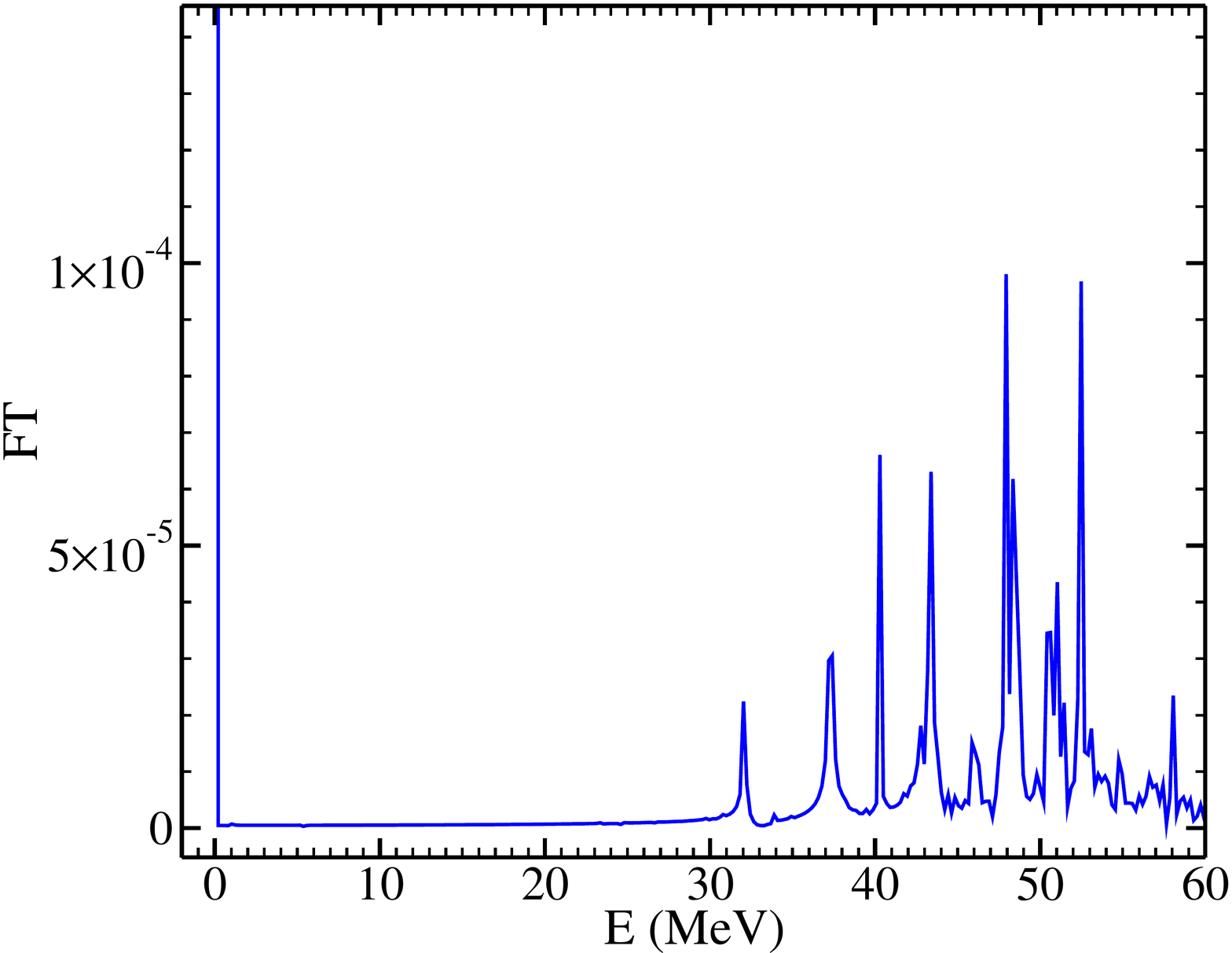}
\end{center}
\caption{(Color online) Fourier transform of a randomly-selected cluster amplitude
  $s_i^a$ or $s_{ij}^{ab}$ for $^{16}$O. Energies are measured
  relative to the ground-state energy.}\label{fig:o16ft}
\end{figure}

In coupled-cluster theory, the standard approach to excited
states~\cite{KocJor90,StaBar93} is the solution of the right
eigenvalue problem~(\ref{eq:eom}).  Alternatively, we can obtain a
spectrum of excited-state energies from a Fourier transformation of
any of the time-evolved CCSD amplitudes $s_i^a$ and $s_{ij}^{ab}$,
utilizing the solution of Eq.~(\ref{tibasis}).  This approach involves
first solving the time-independent CCSD equations, Eqs.~(\ref{eq:tia})
and (\ref{eq:tijab}).  Then, we perturb the resulting cluster
amplitudes by a small random contribution of the order $\approx$~0.001
and solve Eqs.~(\ref{eq:tdccs12}) and (\ref{eq:tdccs22}).  During the
iterations, we record the time evolution of a single,
randomly-selected amplitude $s_i^a$ or $s_{ij}^{ab}$.  Results for the
Fourier transform of the CCSD amplitudes for $^4$He and $^{16}$O are
shown in Fig.~\ref{fig:he4ft} and Fig.~\ref{fig:o16ft}, respectively.
The length of the time evolution is $20000$~fm/$c$ for $^4$He and
$6000$~fm/$c$ for $^{16}$O, and the step size is 1~fm/$c$.
Corresponding to the length of the time evolution, the energy
uncertainty in the resulting Fourier spectra is about $\delta E\approx
0.06$~MeV for $^4$He and $\delta E\approx 0.2$~MeV for $^{16}$O. In
each plot (the excitation energies are measured relative to the ground
state) the giant peaks at $E=0$ correspond to the ground-state of each
system.  Note that several of the excited-state energies shown in
Figs.~\ref{fig:he4ft} and \ref{fig:o16ft} are not associated with
physical states of $^4$He or $^{16}$O, respectively but arise from the
limitation of a small model space, and the incomplete separation of
the center-of-mass motion~\cite{HagPapEA09}.  In Tables~\ref{tb:he4}
and~\ref{tb:o16}, for $^4$He and $^{16}$O, respectively, we compare
the values associated with four selected peaks to the values computed
using equation-of-motion CCSD.  Note that all of the peaks in
Figs.~\ref{fig:he4ft} and \ref{fig:o16ft} are associated with energies
computed with this method.  The agreement is good, and deviations are
within the uncertainties related to the finite time evolution.

\begin{table}[h]
  \caption{Comparison of select excited-state energies obtained via time-dependent CCSD (TD-CCSD) and equation-of-motion CCSD (EOM-CCSD) for $^{4}$He.  Energies are given in units of MeV.}
\label{tb:he4}
\centering
\begin{tabular}{lrrrr}
\hline\hline
& ~~TD-CCSD & ~~EOM-CCSD \\
\hline\hline
$E_1$ &  ~3.04 & ~3.02 \\
$E_2$ &  29.45 & 29.47 \\
$E_3$ &  35.52 & 35.53 \\
$E_4$ &  41.10 & 41.11 \\
\hline
\end{tabular}
\end{table}

\begin{table}[h]
  \caption{Comparison of select excited-state energies obtained via time-dependent CCSD (TD-CCSD) and equation-of-motion CCSD (EOM-CCSD) for $^{16}$O.  Energies are given in units of MeV.}
\label{tb:o16}
\centering
\begin{tabular}{lrrrr}
\hline\hline
& ~~TD-CCSD & ~~EOM-CCSD \\
\hline\hline
$E_1$ &  32.20 & 32.29 \\
$E_2$ &  33.85 & 33.91 \\
$E_3$ &  37.40 & 37.60 \\
$E_4$ &  48.15 & 48.12 \\
\hline
\end{tabular}
\end{table}

Note that TDCC calculation of excited states via Fourier
transformation requires much more computational cycles than the
standard approach. The latter requires us to solve the eigenvalue
problem~(\ref{eq:eom}) once, while the TDCC method requires one
computation of $\overline{H}_1$ per time step. The real advantage of
the time-dependent coupled-cluster method lies -- of course -- in the
study of time-dependent Hamiltonians.

\subsection{\label{sec:lip}Excitation energies of interacting Lipkin systems}

Hoodbhoy and Negele studied the dynamics of two interacting,
14-particle Lipkin models~\cite{Lipkin} with the TDCC
method~\cite{HooNeg78}. In this Subsection, we compare our
calculations to the earlier results.

The $N$-particle Lipkin model consists of two $N$-fold degenerate
levels separated by an energy $\varepsilon$.  In the reference state, all
$N$ particles occupy the lower level.  The particles interact via the
two-body Hamiltonian
\begin{eqnarray}
  H&=&\frac{\varepsilon}{2}\sum_{p\sigma}\sigma a_{p\sigma}^{\dag}a_{p\sigma}+\frac{V}{2}\sum_{pq\sigma}a_{p\sigma}^{\dag}a_{q\sigma}^{\dag}a_{q-\sigma}a_{p-\sigma} \ .
\end{eqnarray}
Here $p,q=1,\ldots,N$ are particle labels, $\sigma=\pm 1$ denotes the
upper and lower level, respectively, and $V$ is the strength of the
two-body interaction. For weak interactions $V(N-1)< \varepsilon$, the
reference state is $\prod_{p=1}^N a^\dagger_{p-}|0\rangle$, and the
model is in its ``symmetric'' phase.  At $V(N-1)\approx\varepsilon$
the model makes a transition to a ``deformed'' phase, and a
Hartree-Fock reference state should be chosen for $V(N-1)>\varepsilon$.

As in Ref.~\cite{HooNeg78} we consider two Lipkin models ($N=14$ each
with Hamiltonians $H_1$ and $H_2$, respectively) that are
noninteracting for $t<0$ and start to interact at $t=0$ via 
\begin{eqnarray}
H_{\rm int}=V\sum_{p_1p_2\sigma}a_{p_1\sigma}^{\dag}a_{p_2\sigma}^{\dag}a_{p_2-\sigma}a_{p_1-\sigma} \ .
\label{eq:h12}
\end{eqnarray}
The observable of interest is the energy difference
\begin{eqnarray}
\Delta E(t)&\equiv&\frac{1}{2}\Big(\Bra{\Psi}H_1+H_2\Ket{\Psi}(t)\label{eq:deltae}\\
&&~-\Bra{\Psi}H_1+H_2\Ket{\Psi}(0)\Big)\nonumber 
\end{eqnarray}
of the subsystems. 

\begin{figure}[t]
\begin{center}
\includegraphics[keepaspectratio=true,scale=0.300]{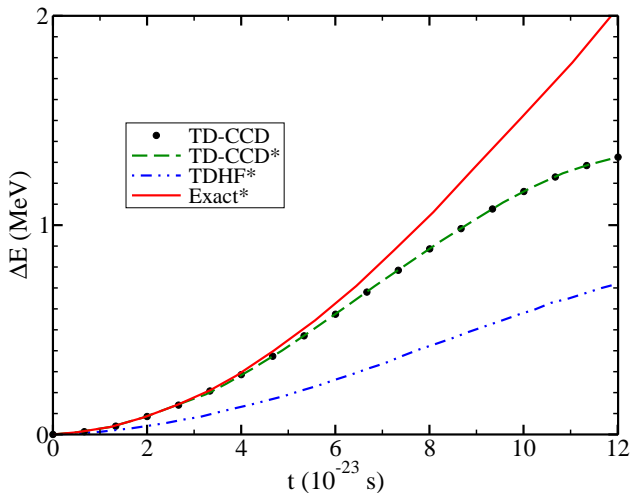}
\end{center}
\caption{(Color online) Time-dependent coupled-cluster doubles
  (TD-CCD) result for the excitation energy as a function of time for
  two interacting, 14-particle Lipkin systems.  Hoodbhoy and Negele's
  results for TD-CCD, time-dependent Hartree-Fock (TDHF), and the
  exact calculation are noted with an asterisk and were taken from
  Ref.~\cite{HooNeg78}.}
\label{fig:hn}
\end{figure}

In Hoodbhoy and Negele's calculation only 2$p$-2$h$ excitations were
considered (setting $S_1=0$), and Eq.~(\ref{eq:tdccs22}) is solved.
The calculation of expectation values was performed in a Hermitian
approach based on the density matrices
\begin{eqnarray}
({\rho_{qp}})_N&=&\Bra{\Phi}e^{S^{\dag}}\{p^{\dag}q\}e^{S}\Ket{\Phi}_C , \label{eq:1den1}\\
({\rho_{rspq}})_N&=&\Bra{\Phi}e^{S^{\dag}}\{p^{\dag}q^{\dag}sr\}e^{S}\Ket{\Phi}_C\label{eq:1den2} .
\end{eqnarray}
Here the subscripts $N$ and the curly brackets indicate the
normal-ordering of the creation and annihilation operators. The
subscript $C$ indicates that the disconnected diagrams have been
factored from the expressions.  Whereas these expressions can easily
be expanded in terms of products of the cluster amplitudes $s_i^a$ and
$s_{ij}^{ab}$, the expansions are non-terminating due to the
exponential nature of the excitation operators. Thus they must be
truncated at some level, and we choose a truncation at the terms
quadratic in the $s_{ij}^{ab}$ amplitudes. Setting $\varepsilon=1$ and
$V=0.357$ for strong coupling, we computed the excitation energy of
each system as a function of time for $12\times 10^{-23}$~s, making 90
time steps of size 0.4~fm/$c$.  Our result for the excitation energy
and that obtained by Hoodbhoy and Negele~\cite{HooNeg78} are shown in
Fig.~\ref{fig:hn}.  The results are indistinguishable.
Figure~\ref{fig:hn} also shows the exact results and results from
time-dependent Hartree-Fock (both taken from Ref.~\cite{HooNeg78}).

\begin{figure}[t]
\begin{center}
\includegraphics[keepaspectratio=true,scale=0.300]{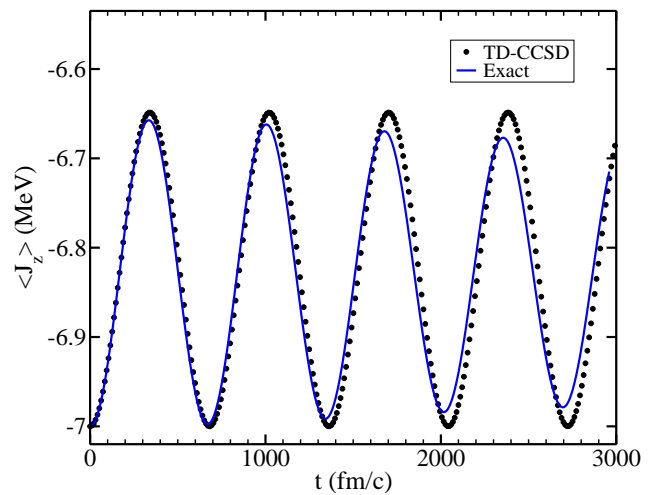}
\end{center}
\caption{(Color online) Time-dependent CCSD evolution of the one-body interaction of
  the 14-particle Lipkin system.  The exact result is also shown.}
\label{fig:jz}
\end{figure}

Let us finally also compute time-dependent expectation values within
the coupled-cluster framework and employ the evolution of the left
de-excitation operator $L$ and the density matrix~(\ref{eq:3den1}) of
the 14-particle Lipkin system.  The parameters of the Lipkin model are
$\varepsilon=1$ and $V=0.04$. Thus, $(N-1)V\approx 0.52 < 1$, and we are
in the symmetric phase of the two-level Lipkin model.  We choose the
initial conditions $S=0=L$ and study the time evolution of the
operator
\begin{eqnarray}
J_z&=&\frac{1}{2}\sum_{p\sigma}\sigma a_{p\sigma}^{\dag}a_{p\sigma} .
\end{eqnarray}
Our result shown in Fig.~\ref{fig:jz} exhibits a good agreement with
the exact solution. The main period and amplitude of the oscillation
is well reproduced but the coupled-cluster calculation misses a fine
modulation of the amplitude. We also performed computations for weaker
interaction strengths $V$ and observed that the agreement between TDCC
theory and the exact solution improves with decreasing values of the
interaction strength.

\section{Conclusions}

We have studied the bi-variational formulation of time-dependent
coupled-cluster theory and illustrated some of its features in
applications to simple models of the atomic nuclei. Our main results
are (i) the explicit demonstration that observables that commute with
the Hamiltonian are conserved under time evolution, (ii) the
explanation of the role of the similarity-transformed Hamiltonian
under real and imaginary time evolution, and (iii) the computation of
energy spectra and time-dependent observables and their comparison
with exact results within simple models. These results open the way
for the description of time-dependent phenomena such as nuclear
collisions within more realistic models and beyond time-dependent
mean-field methods.

\begin{acknowledgments}
  The authors thank G. F. Bertsch, D. J. Dean, S.  Kvaal, and S. Umar
  for discussions. This work has been supported by the U.S.
  Department of Energy under grant Nos. DE-FG02-96ER40975 (Vanderbilt
  University) and DE-FG02-96ER40963 (University of Tennessee) and
  DE-AC05-00OR22725 with UT-Battelle, LLC (Oak Ridge National
  Laboratory).  This research used resources of the Oak Ridge
  Leadership Computing Facility at the Oak Ridge National Laboratory.
\end{acknowledgments}

\appendix

\section{\label{app:tdb}Equations~(\ref{eq:tdcce2}), (\ref{eq:tdccs12}), and (\ref{eq:tdccs22}) in a time-dependent basis}

If the single-particle states are allowed to time-evolve, the
commutator in Eq.~(\ref{eq:hbctdccs}) is nonzero, and the equations of
motion for the cluster amplitudes become
\begin{eqnarray}
\Bra{\Phi}\overline{H}\Ket{\Phi}&=&
i\hbar\Big[\dot{s}_0+\sum_k\braket{k|\dot{k}}+\sum_{kc}s_k^c\braket{k|\dot{c}}\Big] \ ,
\label{eq:tdcce2f}\\
\Bra{\Phi_{i}^{a}}\overline{H}\Ket{\Phi}&=&
i\hbar\Big[\dot{s}_i^a+\braket{a|\dot{i}}
\label{eq:tdccs12f}\\
&&{}+\sum_c s_i^c\braket{a|\dot{c}}+\sum_ ks_k^a\braket{i|\dot{k}}^*\nonumber\\
&&{}+\frac{1}{2}\sum_{kc}s_k^as_i^c\left(\braket{c|\dot{k}}^*-\braket{k|\dot{c}}\right)
\nonumber\\
&&{}+\sum_{kc}s_{ki}^{ca}\left(\braket{k|\dot{c}} + \braket{c|\dot{k}}^*\right) \Big] \ ,
\nonumber\\
\Bra{\Phi_{ij}^{ab}}\overline{H}\Ket{\Phi}&=&i\hbar\Big[\dot{s}_{ij}^{ab}
+ \sum_c\left(s_{ij}^{cb}\braket{a|\dot{c}}-s_{ij}^{ca}\braket{b|\dot{c}}\right)
\label{eq:tdccs22f}\\
&&{}+\sum_k\left(s_{kj}^{ab}\braket{i|\dot{k}}-s_{ki}^{ab}\braket{j|\dot{k}}\right)\nonumber\\
&&{}+\frac{1}{2}P(ab)\sum_{kc}s_k^as_{ij}^{cb}\left(\braket{c|\dot{k}}^*-\braket{k|\dot{c}}\right)\nonumber\\
&&{}+\frac{1}{2}P(ij)\sum_{kc}s_i^cs_{kj}^{ab}\left(\braket{c|\dot{k}}^*-\braket{k|\dot{c}}\right)\Big] \ .\nonumber
\end{eqnarray}
Here it is clear that the creation and annihilation operators are
time-dependent, and the permutation operator $P$ is defined as
$P(pq)f(p,q)\equiv f(p,q) - f(q,p)$.

\bibliography{ref}

\end{document}